\documentclass{article}

\PassOptionsToPackage{numbers}{natbib}

\usepackage[preprint]{neurips_2022}

\usepackage{amsmath,amsfonts,bm}

\def\eqref#1{equation~\ref{#1}}

\def\1{\bm{1}}

\def\vd{{\bm{d}}}

\def\mC{{\bm{C}}}

\def\mE{{\bm{E}}}

\def\mH{{\bm{H}}}

\def\mO{{\bm{O}}}

\def\mS{{\bm{S}}}

\def\mW{{\bm{W}}}

\DeclareMathAlphabet{\mathsfit}{\encodingdefault}{\sfdefault}{m}{sl}
\SetMathAlphabet{\mathsfit}{bold}{\encodingdefault}{\sfdefault}{bx}{n}

\usepackage[utf8]{inputenc} 
\usepackage[T1]{fontenc}    
\usepackage{booktabs}       
\usepackage{amsfonts}       
\usepackage{nicefrac}       
\usepackage{microtype}      
\usepackage{amsmath}
\usepackage{graphicx}
\usepackage{tipa}
\usepackage{multirow}
\usepackage{longtable}

\usepackage{subcaption}
\usepackage{color}
\usepackage[table,dvipsnames]{xcolor}
\usepackage{threeparttable}
\usepackage{diagbox}
\usepackage{enumitem}
\usepackage{tablefootnote}
\usepackage{hyperref}
\usepackage{url}            
\usepackage{xurl}
\usepackage{breakurl}
\usepackage{algorithm}
\usepackage{algpseudocode}
\usepackage{wrapfig}

\usepackage{amsthm}
\theoremstyle{definition}
\newtheorem{definition}{Definition}

\usepackage[edges]{forest}
\usepackage{tikz}
\usepackage{tikz-qtree}
\usetikzlibrary{fadings}
\usetikzlibrary{mindmap,trees}
\usetikzlibrary{arrows,automata,shapes,positioning,shadows,trees}
\usetikzlibrary{shapes,snakes,shadows}
\usetikzlibrary{shapes.arrows}
\usetikzlibrary{calc,shapes, positioning}
\usetikzlibrary{decorations.text}
\usetikzlibrary{matrix,chains,positioning,decorations.pathreplacing,arrows}
\usetikzlibrary{bayesnet}
\usetikzlibrary{shadows.blur}
\usetikzlibrary{shapes.geometric}

\tikzstyle{edge}=[-latex',draw=black!90,shorten <=1pt,shorten >=1pt]
\tikzstyle{redge}=[latex'-,draw=black!90,shorten <=1pt,shorten >=1pt]
\tikzstyle{dedge}=[latex'-latex',draw=black!90,shorten <=1pt,shorten >=1pt]
\tikzstyle{block}=[draw, text width=5em,align=center,shape=rectangle, rounded corners, , align=center]
\tikzstyle{nobox}=[align=center]

\definecolor{emb}{RGB}{209,228,252}
\definecolor{hidden-blue}{RGB}{194,232,247}
\definecolor{hidden-orange}{RGB}{243,202,120}
\definecolor{hidden-yellow}{RGB}{242,244,193}
\definecolor{output-purple}{RGB}{219,203,231}
\definecolor{output-green}{RGB}{204,231,207}
\definecolor{hiddendraw}{RGB}{205, 44, 36}

\tikzstyle{mybox}=[
    rectangle,
    draw=hiddendraw,
    rounded corners,
    text opacity=1,
    minimum height=1.5em,
    minimum width=5em,
    inner sep=2pt,
    align=center,
    fill opacity=.2,
    ]

\tikzstyle{emb-purple}=[
    rectangle,
    draw=output-purple!50!purple,
    fill=output-purple,
    text opacity=1,
    minimum height=1.5em,
    minimum width=1.5em,
    inner sep=0pt,
    align=center,
    fill opacity=.9,
    ]

\tikzstyle{emb-blue}=[
    rectangle,
    draw=emb!50!blue,
    fill=emb,
    text opacity=1,
    minimum height=1.5em,
    minimum width=1.5em,
    inner sep=0pt,
    align=center,
    fill opacity=.9,
]

\definecolor{colone}{RGB}{178, 34, 34}
\definecolor{coltwo}{RGB}{106, 90, 205}
\definecolor{colthree}{RGB}{255, 250, 205}
\definecolor{colfour}{RGB}{0, 139, 69}
\definecolor{colfive}{RGB}{245,238,197}
\definecolor{colsix}{RGB}{243,235,179}
\definecolor{colseven}{RGB}{241,231,163}
\def\myname{NaturalSpeech}

\title{\textit{\myname{}}: End-to-End Text to Speech Synthesis with Human-Level Quality}

\author{
Xu Tan\thanks{Equal contribution. Corresponding author: Xu Tan, \texttt{xuta@microsoft.com}}, ~Jiawei Chen\footnotemark[1], ~Haohe Liu\footnotemark[1], ~Jian Cong, Chen Zhang, Yanqing Liu, Xi Wang  \\
\textbf{Yichong Leng, Yuanhao Yi, Lei He, Frank Soong} \\  
\textbf{Tao Qin, Sheng Zhao, Tie-Yan Liu}  \\
\\
Microsoft Research Asia \& Microsoft Azure Speech \\
}

\begin{document}

\maketitle

\begin{abstract}
Text to speech (TTS) has made rapid progress in both academia and industry in recent years. Some questions naturally arise that whether a TTS system can achieve human-level quality, how to define/judge that quality and how to achieve it. In this paper, we answer these questions by first defining the human-level quality based on the statistical significance of subjective measure and introducing appropriate guidelines to judge it, and then developing a TTS system called \myname{} that achieves human-level quality on a benchmark dataset. Specifically, we leverage a variational autoencoder (VAE) for end-to-end text to waveform generation, with several key modules to enhance the capacity of the prior from text and reduce the complexity of the posterior from speech, including phoneme pre-training, differentiable duration modeling, bidirectional prior/posterior modeling, and a memory mechanism in VAE. Experiment evaluations on popular LJSpeech dataset show that our proposed \myname{} achieves $-0.01$ CMOS (comparative mean opinion score) to human recordings at the sentence level, with Wilcoxon signed rank test at p-level $p \gg 0.05$, which demonstrates no statistically significant difference from human recordings for the first time on this dataset. 

\end{abstract}

\section{Introduction}

Text to speech (TTS) aims at synthesizing intelligible and natural speech from text~\cite{tan2021survey}, and has made rapid progress in recent years due to the development of deep learning. Neural network based TTS has evolved from CNN/RNN-based models~\cite{oord2016wavenet,shen2018natural,wang2017tacotron,arik2017deep,gibiansky2017deep,ping2018deep,tachibana2018efficiently} to Transformer-based models~\cite{li2019neural,ren2019fastspeech,liu2021delightfultts}, from basic generative models (autoregressive)~\cite{oord2016wavenet,shen2018natural,li2019neural} to more powerful models (VAE, GAN, flow, diffusion)~\cite{prenger2019waveglow,kim2020glow,popov2021grad,kim2021conditional}, from cascaded acoustic models/vocoders~\cite{oord2016wavenet,wang2017tacotron,shen2018natural,ren2019fastspeech,kalchbrenner2018efficient,kong2020hifi} to fully end-to-end models~\cite{ren2021fastspeech,donahue2020end,kim2021conditional}.

Building TTS systems with human-level quality has always been the dream of the practitioners in speech synthesis. While current TTS systems achieve high voice quality, they still have quality gap compared with human recordings. To pursue this goal, several questions need to be answered: 1) how to define human-level quality in text to speech synthesis? 2) how to judge whether a TTS system has achieved human-level quality or not? 3) how to build a TTS system to achieve human-level quality? In this paper, we conduct a comprehensive study on these problems in TTS. We first give a formal definition on human-level quality in TTS based on a statistical and measurable way (see Definition~\ref{def_onpar}). Then we introduce some guidelines to judge whether a TTS system has achieved human-level quality with a hypothesis test. Using this judge method, we found several previous TTS systems have not achieved it (see Table~\ref{tab_mos_cmos_judge_onpar}).

In this paper, we further develop a fully end-to-end text to waveform generation system called \myname{} to bridge the quality gap to recordings and achieve human-level quality. Specifically, inspired by image/video/waveform generation~\cite{van2017neural,ramesh2021zero,kim2021conditional}, we leverage variational autoencoder (VAE)~\cite{kingma2013auto} to compress the high-dimensional speech ($x$) into continuous frame-level representations (denoted as posterior $q(z|x)$), which are used to reconstruct the waveform (denoted as $p(x|z)$). The corresponding prior (denoted as $p(z|y)$) is obtained from the text sequence $y$. Considering the posterior from speech is more complicated than the prior from text, we design several modules (see Figure~\ref{fig_model_overall}) to match the posterior and prior as close to each other as possible, to enable text to speech synthesis through $p(z|y) \rightarrow p(x|z)$:

\begin{itemize}[leftmargin=*]
\item We leverage large-scale pre-training on the phoneme encoder to extract better representations from phoneme sequence (Section~\ref{sec_phoneme_pretrain}).  

\item We leverage a fully differentiable durator\footnote{Since duration is very important in TTS, especially in non-autoregressive TTS, we name the module related to duration modeling as \textit{durator}, including but not limited to the functionalities of duration prediction and hidden expansion. It is common to come up with new term to revolutionize the concept in speech community, such as \textit{vocoder}, \textit{cepstrum}.} that consists of a duration predictor and an upsampling layer to improve the duration modeling (Section~\ref{sec_diff_durator}). 

\item We design a bidirectional prior/posterior module based on flow models~\cite{dinh2014nice,kingma2016improved,kingma2018glow} to further enhance the prior $p(z|y)$ and reduce the complexity of posterior $q(z|x)$ (Section~\ref{sec_bidirect_flow}).

\item We propose a memory based VAE to reduce the complexity of the posterior needed to reconstruct waveform (Section~\ref{sec_memory_vae}). 

\end{itemize}

Compared to previous TTS systems, \myname{} has several advantages: 1) Reduce training-inference mismatch. In previous cascaded acoustic model/vocoder pipeline~\citep{kim2020glow,ren2021fastspeech,popov2021grad} and explicit duration prediction~\citep{kim2020glow,kim2021conditional,ren2021fastspeech}, both mel-spectrogram and duration suffer from training-inference mismatch since ground-truth values are used in training the vocoder and mel-spectrogram decoder while predicted values are used in inference. Our fully end-to-end text to waveform generation and differentiable durator can avoid the training-inference mismatch. 2) Alleviate one-to-many mapping problem. One text sequence can correspond to multiple speech utterances with different variation information (e.g., pitch, duration, speed, pause, prosody, etc). Previous works only using variance adaptor~\citep{ren2021fastspeech,liu2021delightfultts} to predict pitch/duration cannot well handle the one-to-many mapping problem. Our memory based VAE and bidirectional prior/posterior can reduce the complexity of posterior and enhance the prior, which helps relieve the one-to-many mapping problem. 3) Improve representation capacity. Previous models are not powerful enough to extract good representations from phoneme sequence~\citep{kim2020glow,kim2021conditional,popov2021grad} and learn complicated data distribution in speech~\citep{ren2021fastspeech}. Our large-scale phoneme pre-training and powerful generative models such as flow and VAE can learn better text representations and speech data distributions.

We conduct experimental evaluations on the widely adopted LJSpeech dataset~\cite{ljspeech17} to measure the voice quality of our \myname{} system. Based on the proposed judgement guidelines, \myname{} achieves similar quality with human recordings in terms of MOS (mean opinion score) and CMOS (comparative MOS). Specifically, the speech generated by \myname{} achieves $-0.01$ CMOS compared to recordings, with p-level $p \gg 0.05$ under Wilcoxon signed rank test, which demonstrates that \myname{} can generate speech with no statistically significant difference from recordings.

\section{Definition and Judgement of Human-Level Quality in TTS}
\label{sec_def_judge}

In this section, we introduce the formal definition of human-level quality in text to speech synthesis, and describe how to judge whether a TTS system achieves human-level quality or not. 

\subsection{Definition of Human-Level Quality}
We define human-level quality in a statistical and measurable way.
\begin{definition}
\textit{If there is no statistically significant difference between the quality scores of the speech generated by a TTS system and the quality scores of the corresponding human recordings on a test set, then this TTS system achieves human-level quality on this test set.}
\label{def_onpar}
\end{definition}

Note that by claiming a TTS system achieves human-level quality on a test set, we do not mean that a TTS system can surpass or replace human, but the quality of this TTS system is statistically indistinguishable from human recordings on this test set.

\subsection{Judgement of Human-Level Quality}

\paragraph{Judgement Guideline} While there are some objective metrics to measure the quality gap between the generated speech and human recordings, such as PESQ~\citep{rix2001perceptual}, STOI~\citep{taal2011algorithm}, SI-SDR~\citep{le2019sdr}, they are not reliable to measure the perception quality in TTS. Therefore, we use subjective evaluation to measure the voice quality. Previous works usually use mean opinion score (MOS) with $5$ points (from $1$ to $5$) to compare the generated speech with recordings. However, MOS is not sensitive enough to the difference in voice quality since the judge simply rates the quality of each sentence alone from the two systems with no paired comparison. Thus, we choose comparative mean opinion score (CMOS) with 7 points (from $-3$ to $3$) as the evaluation metric, where each judge measures the voice quality by comparing samples from two systems head by head. We further conduct Wilcoxon signed rank test~\citep{wilcoxon1992individual} to measure whether the two systems are significantly different or not in terms of CMOS evaluation. 

Therefore, we list the judgement guidelines of human-level quality as follows: 1) Each utterance from TTS system and human recordings should be listened and compared side-by-side by more than 20 judges, who should be native language speakers. At least 50 test utterances from each system should be used in the judgement.
2) The speech generated by TTS system has no statistically significant difference from human recordings, if and only if the average CMOS is close to $0$ and the p-level of Wilcoxon signed rank test satisfies $p>0.05$. 

\paragraph{Judgement of Previous TTS Systems} 
Based on these guidelines, we test whether current TTS systems can achieve human-level quality or not on the LJSpeech dataset. The systems we study include: 1) FastSpeech 2~\citep{ren2021fastspeech} + HiFiGAN~\citep{kong2020hifi}, 2) Glow-TTS~\citep{kim2020glow} + HiFiGAN~\citep{kong2020hifi}, 3) Grad-TTS~\citep{popov2021grad} + HiFiGAN~\citep{kong2020hifi}, 4) VITS~\citep{kim2021conditional}. We re-produce the results of all these systems by our own, which can match or even beat the quality in their original papers (note that the HiFiGAN vocoder is fine-tuned on the predicted mel-spectrograms for better synthesis quality). We use 50 test utterances, each with 20 judges for MOS and CMOS evaluation. As shown in Table~\ref{tab_mos_cmos_judge_onpar}, although the current TTS systems can achieve close MOS with recordings, they have a large CMOS gap to recordings, with Wilcoxon signed rank test at p-level $p \ll 0.05$, which shows statistically significant difference from human recordings. We further study where the quality gap comes from by analyzing each component in one of the above TTS systems in Appendix~\ref{appendix_study}.

\begin{table}[h!]
\small
\caption{The MOS and CMOS comparisons between previous TTS systems and human recordings. Note that the Wilcoxon p-value in MOS is conducted using Wilcoxon rank sum test~\citep{wilcoxon1992individual}, instead of the Wilcoxon signed rank test in CMOS, due to no paired comparison in MOS evaluation. For Grad-TTS, we use 1000 steps for inference.}
\centering
\begin{tabular}{ l | c c | c c}
    \toprule
	    System & MOS & Wilcoxon p-value & CMOS & Wilcoxon p-value \\
	    \midrule
	    Human Recordings  & $4.52\pm0.11$  & - & $0$ & - \\
	    \midrule
	    FastSpeech 2~\citep{ren2021fastspeech} + HiFiGAN~\citep{kong2020hifi} & $4.32\pm0.10$ & $1.0\text{e-}05$ & $-0.30$ & $5.1\text{e-}20$ \\
	    Glow-TTS~\citep{kim2020glow} + HiFiGAN~\citep{kong2020hifi} & $4.33\pm0.10$ & $1.3\text{e-}06$ & $-0.23$ & $8.7\text{e-}17$\\
	    Grad-TTS~\citep{popov2021grad} + HiFiGAN~\citep{kong2020hifi}  & $4.37\pm0.10$  & $0.0127$ & $-0.23$ & $1.2\text{e-}11$ \\
	    VITS~\citep{kim2021conditional} &  $4.49\pm0.10$ & $0.2429$ & $-0.19$ & $2.9\text{e-}04$ \\
	    \bottomrule
	\end{tabular}
	\vspace{0.3cm}
	\label{tab_mos_cmos_judge_onpar}
\end{table}

\section{Description of \myname{} System}
\label{sec_method}

To bridge the quality gap to human recordings, we develop \myname{}, a fully end-to-end text to waveform generation model. We first describe the design principle of our system (Section~\ref{sec_deign_principle}), and then introduce each module of this system (Section~\ref{sec_phoneme_pretrain}-\ref{sec_memory_vae}) and training/inference pipeline (Section~\ref{sec_train_infer}), and finally explain why our system can bridge the quality gap to human recordings (Section~\ref{sec_why_can_close}). 

\subsection{Design Principle}
\label{sec_deign_principle}

Inspired by image/video generation~\cite{ramesh2021zero,ding2021cogview,wu2021nuwa,yan2021videogpt,rakhimov2020latent} that uses VQ-VAE~\cite{van2017neural,razavi2019generating,esser2021taming} to compress high-dimensional image into low-dimensional representations to ease the generation, we leverage VAE~\cite{kingma2013auto} to compress high-dimensional speech $x$ into frame-level representations $z$ (i.e., $z$ is sampled from posterior distribution $q(z|x)$), which are used to reconstruct the waveform (denoted as $p(x|z)$). In general formulation of VAE, the prior $p(z)$ is chosen to be standard isotropic multivariate Gaussian. To enable conditional waveform generation from input text in TTS, we predict $z$ from phoneme sequence $y$, i.e., $z$ is sampled from predicted prior distribution $p(z|y)$. We jointly optimize the VAE and the prior prediction with gradients propogating to both $q(z|x)$ and $p(z|y)$. Derived from the evidence lower bound~\cite{kingma2013auto}, the loss function consists of a waveform reconstruction loss $-\log p(x|z)$ and a Kullback-Leibler divergence loss between the posterior $q(z|x)$ and the prior $p(z|y)$, i.e., $KL[q(z|x)||p(z|y)]$.

\begin{figure} [t!]
\centering
    \includegraphics[page=1,width=0.65\columnwidth,trim=4.5cm 1.9cm 4.6cm 1.9cm,clip=true]{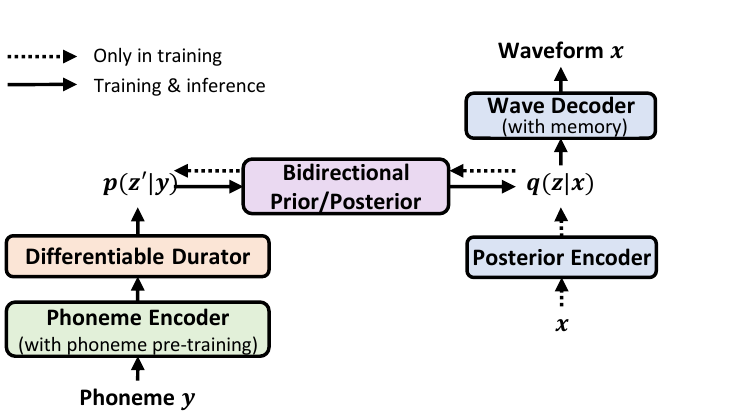}
    \caption{System overview of \myname{}.}
  \label{fig_model_overall}
\vspace{-0.3cm}
\end{figure}

Considering the posterior from speech is more complicated than the prior from text, to match them as close as possible to enable text to waveform generation, we design several modules to simplify the posterior and to enhance the prior, as shown in Figure~\ref{fig_model_overall}. First, to learn a good representations of phoneme sequence for better prior prediction, we pre-train a phoneme encoder on a large-scale text corpus using masked language modeling on phoneme sequence (Section~\ref{sec_phoneme_pretrain}). Second, since the posterior is at the frame level while the phoneme prior is at the phoneme level, we need to expand the phoneme prior according to its duration to bridge the length difference. We leverage a differentiable durator to improve duration modeling (Section~\ref{sec_diff_durator}). Third, we design a bidirectional prior/posterior module to enhance the prior or simplify the posterior (Section~\ref{sec_bidirect_flow}). Fourth, we propose a memory based VAE that leverages a memory bank through Q-K-V attention~\citep{vaswani2017attention} to reduce the complexity of posterior needed to reconstruct the waveform (Section~\ref{sec_memory_vae}).

\subsection{Phoneme Encoder}
\label{sec_phoneme_pretrain}
The phoneme encoder $\theta_{\text{pho}}$ takes a phoneme sequence $y$ as input and outputs a phoneme hidden sequence. To enhance the representation capability of the phoneme encoder, we conduct large-scale phoneme pre-training. Previous works~\citep{xiao2020improving} conduct pre-training in character/word level and apply the pre-trained model to phoneme encoder, which will cause inconsistency, and the works~\citep{jia2021png} directly using phoneme pre-training will suffer from limited capacity due to too small size of phoneme vocabulary. To avoid these issues, we leverage mixed-phoneme pre-training~\citep{zhang2022mixed}, which uses both phoneme and sup-phoneme (adjacent phonemes merged together) as the input of the model, as shown in Figure~\ref{fig_model_pho}. When using masked language modeling~\cite{devlin2018bert}, we randomly mask some sup-phoneme tokens and their corresponding phoneme tokens and predict the masked phoneme and sup-phoneme at the same time. After mixed phoneme pre-training, we use the pre-trained model to initialize the phoneme encoder of our TTS system.

\subsection{Differentiable Durator}
\label{sec_diff_durator}

\begin{figure} [t!]
\centering
    \begin{subfigure}[b]{0.32\textwidth}
    \includegraphics[page=3,width=1.0\columnwidth,trim=6cm 3.0cm 6.5cm 3.1cm,clip=true]{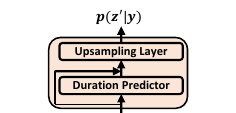} 
    \caption{Differentiable durator.}
    \label{fig_model_dur}
    \end{subfigure}
    \begin{subfigure}[b]{0.67\textwidth}
    \includegraphics[page=2,width=0.8\columnwidth,trim=4.5cm 3.3cm 5.0cm 3.3cm,clip=true]{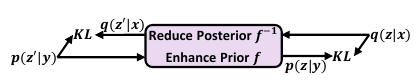}
    \caption{Bidirectional prior/posterior.}
    \label{fig_model_bpp}
    \end{subfigure}\hspace{-0.7cm}
	\begin{subfigure}[b]{0.54\textwidth}
    \includegraphics[page=6,width=1\columnwidth,trim=4.5cm 2.7cm 5cm 2.55cm,clip=true]{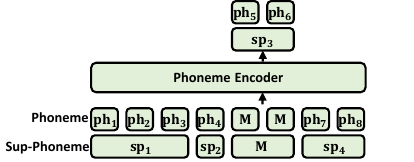}
    \caption{Phoneme pre-training.}
    \label{fig_model_pho}
    \end{subfigure}\hspace{0.0cm}
    \begin{subfigure}[b]{0.39\textwidth}
    \includegraphics[page=5,width=1\columnwidth,trim=6cm 2.5cm 5.5cm 2.3cm,clip=true]{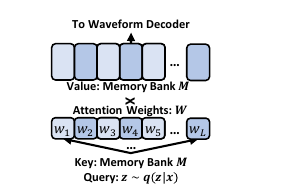} 
    \caption{Memory mechanism in VAE.}
    \label{fig_model_mem}
    \end{subfigure}
\caption{The designed modules in \myname{}.}
\label{fig_each_design}
\vspace{-0.2cm}
\end{figure}

The differentiable durator $\theta_{\text{dur}}$ takes a phoneme hidden sequence as input, and outputs a sequence of prior distribution at the frame level, as shown in Figure~\ref{fig_model_dur}. We denote the prior distribution as $p(z'|y; \theta_{\text{pho}}, \theta_{\text{dur}}) = p(z'|y; \theta_{\text{pri}})$, where $\theta_{\text{pri}} = [\theta_{\text{pho}}, \theta_{\text{dur}}]$. The differentiable durator $\theta_{\text{dur}}$ consists of several modules: 1) a duration predictor that builds upon the phoneme encoder to predict the duration for each phoneme, 2) a learnable upsampling layer that leverages the predicted duration to learn a projection matrix to extend the phoneme hidden sequence from phoneme level to frame level in a differentiable way~\citep{elias2021parallel}, and 3) two additional linear layers on the expanded hidden sequence to calculate the mean and variance of the prior distribution $p(z'|y; \theta_{\text{pri}})$. The detailed formulation of differentiable durator is in Appendix~\ref{appendix_durator}.
We optimize the duration prediction, learnable upsampling layer, and mean/variance linear layers together with the TTS model in a fully differentiable way, which can reduce the training-inference mismatch in previous duration prediction (ground-truth duration is used in training while predicted duration is used in inference)~\citep{kim2020glow,kim2021conditional,ren2021fastspeech} and better use duration in a soft and flexible way instead of a hard expansion, hence the side-effect of inaccurate duration prediction is mitigated.

\subsection{Bidirectional Prior/Posterior}
\label{sec_bidirect_flow}

As shown in Figure~\ref{fig_model_bpp}, we design a bidirectional prior/posterior module to enhance the capacity of the prior $p(z'|y; \theta_{\text{pri}})$ or to reduce the complexity of the posterior $q(z|x;\phi)$ where $\phi$ is the posterior encoder, since there is information gap between the posterior obtained from speech sequence and the prior obtained from phoneme sequence. We choose a flow model~\cite{dinh2014nice,rezende2015variational,kingma2016improved,kingma2018glow} as the bidirectional prior/posterior module (denoted as $\theta_{\text{bpp}}$) since it is easy to optimize and has a nice property of invertibility. 

\paragraph{Reduce Posterior $q(z|x;\phi)$ with Backward Mapping $f^{-1}$}
The bidirectional prior/posterior module can reduce the complexity of posterior from $q(z|x;\phi)$ to $q(z'|x;\phi,\theta_{\text{bpp}})$ through the backward mapping $f^{-1}(z;\theta_{\text{bpp}})$, i.e., for $z \sim q(z|x;\phi)$, $z' = f^{-1}(z;\theta_{\text{bpp}}) \sim q(z'|x;\phi,\theta_{\text{bpp}})$. The objective is to match the simplified posterior $q(z'|x;\phi,\theta_{\text{bpp}})$ to the prior $p(z'|y; \theta_{\text{pri}})$ by using the KL divergence loss as follows:
\begin{equation}
\begin{aligned}
& \mathcal{L}_{\text{bwd}}(\phi,\theta_{\text{bpp}},\theta_{\text{pri}}) =  KL[q(z'|x;\phi,\theta_{\text{bpp}})||p(z'|y; \theta_{\text{pri}}))] = \int q(z'|x;\phi,\theta_{\text{bpp}}) ~\cdot~ \log \frac{q(z'|x;\phi,\theta_{\text{bpp}})}{p(z'|y; \theta_{\text{pri}})} dz' \\
&= \int q(z|x;\phi) |\det \frac{\partial f^{-1}(z; \theta_{\text{bpp}})}{\partial  z} |^{-1} ~\cdot~ \log \frac{q(z|x;\phi)|\det \frac{\partial  f^{-1}(z; \theta_{\text{bpp}})}{\partial z}|^{-1}} {p(f^{-1}(z;\theta_{\text{bpp}})|y; \theta_{\text{pri}})} ~\cdot~ |\det \frac{\partial f^{-1}(z; \theta_{\text{bpp}})}{\partial z}| dz \\ 
&= \int q(z|x;\phi) ~\cdot~ \log \frac{q(z|x;\phi)} {p(f^{-1}(z;\theta_{\text{bpp}})|y; \theta_{\text{pri}}) |\det \frac{\partial  f^{-1}(z; \theta_{\text{bpp}})}{\partial z}|} dz \\
& = \mathbb{E}_{z\sim q(z|x;\phi)} (\log q(z|x;\phi) - \log (p(f^{-1}(z;\theta_{\text{bpp}})|y; \theta_{\text{pri}}) | \det \frac{\partial f^{-1}(z; \theta_{\text{bpp}})}{\partial z} |),
\end{aligned}
\label{eqa_kl_backward}
\end{equation}
where the third equality (the second line) in Equation~\ref{eqa_kl_backward} is obtained via the change of variables: $dz' = |\det \frac{\partial f^{-1}(z; \theta_{\text{bpp}})}{\partial z}| dz$, and $q(z'|x; \phi,\theta_{\text{bpp}}) = q(z|x;\phi)|\det \frac{\partial f(z'; \theta_{\text{bpp}})}{\partial z'}| =  q(z|x;\phi)|\det \frac{\partial  f^{-1}(z; \theta_{\text{bpp}})}{\partial z}|^{-1}$ according to inverse function theorem.

\paragraph{Enhance Prior $p(z'|y; \theta_{\text{pri}})$ with Forward Mapping $f$}
The bidirectional prior/posterior module can enhance the capacity of prior from $p(z'|y; \theta_{\text{pri}})$ to $p(z|y; \theta_{\text{pri}},\theta_{\text{bpp}})$ through the forward mapping $f(z';\theta_{\text{bpp}})$, i.e., for $z' \sim p(z'|y; \theta_{\text{pri}})$, $z = f(z';\theta_{\text{bpp}}) \sim p(z|y; \theta_{\text{pri}}, \theta_{\text{bpp}})$. The objective is to match the enhanced prior $p(z|y; \theta_{\text{pri}},\theta_{\text{bpp}})$ to the posterior $q(z|x;\phi)$ using the KL divergence loss as follows:
\begin{equation}
\begin{aligned}
& \mathcal{L}_{\text{fwd}}(\phi,\theta_{\text{bpp}}, \theta_{\text{pri}}) = KL[p(z|y; \theta_{\text{pri}},\theta_{\text{bpp}})||q(z|x;\phi)] = \int p(z|y; \theta_{\text{pri}},\theta_{\text{bpp}}) ~\cdot~ \log \frac{p(z|y; \theta_{\text{pri}},\theta_{\text{bpp}})}{q(z|x;\phi)} dz \\
&= \int p(z'|y;\theta_{\text{pri}})|\det \frac{\partial f(z'; \theta_{\text{bpp}})}{\partial z'}|^{-1} ~\cdot~ \log \frac{ p(z'|y;\theta_{\text{pri}})|\det \frac{\partial f(z'; \theta_{\text{bpp}})}{\partial z'}|^{-1}} {q(f(z';\theta_{\text{bpp}})|x;\phi)} ~\cdot~ |\det \frac{\partial f(z'; \theta_{\text{bpp}})}{\partial z'}| dz' \\
& = \mathbb{E}_{z' \sim p(z'|y; \theta_{\text{pri}})} (\log p(z'|y; \theta_{\text{pri}}) - \log q(f(z';\theta_{\text{bpp}})|x; \phi) \arrowvert \det \frac{\partial f(z';\theta_{\text{bpp}})}{\partial z'}  \arrowvert),
\end{aligned}
\label{eqa_kl_forward}
\end{equation}
where the third equality (the second line) in Equation~\ref{eqa_kl_forward} is obtained via the change of variables: $dz = |\det \frac{\partial f(z'; \theta_{\text{bpp}})}{\partial z'}| dz'$, and $p(z|y; \theta_{\text{pri}},\theta_{\text{bpp}}) = p(z'|y;\theta_{\text{pri}})|\det \frac{\partial f^{-1}(z; \theta_{\text{bpp}})}{\partial z}| = p(z'|y;\theta_{\text{pri}})|\det \frac{\partial f(z'; \theta_{\text{bpp}})}{\partial z'}|^{-1}$ according to inverse function theorem, similar to that in Equation~\ref{eqa_kl_backward}. 

By using backward and forward loss functions, both directions of the flow model are considered in training, which can reduce the training-inference mismatch in the previous flow models that train in backward direction but infer in forward direction. We also provide another formulation of the bidirectional prior/posterior in Appendix~\ref{appendix_bpp}.

\subsection{VAE with Memory}
\label{sec_memory_vae}

The posterior $q(z|x; \phi)$ in the original VAE model is used to reconstruct the speech waveform, and thus is more complicated than the prior from the phoneme sequence. To further relieve the burden of prior prediction, we simplify the posterior by designing a memory based VAE model. The high-level idea of this design is that instead of directly using $z \sim q(z|x; \phi)$ for waveform reconstruction, we just use $z$ as a query to attend to a memory bank, and use the attention result for waveform reconstruction, as shown in Figure~\ref{fig_model_mem}. In this way, the posterior $z$ is only used to determine the attention weights in the memory bank, and thus is largely simplified. The waveform reconstruction loss based on memory VAE can be formulated as 
\begin{equation}
\begin{aligned}
\mathcal{L}_{\text{rec}}(\phi, \theta_{\text{dec}}) &= - \mathbb{E}_{z \sim q(z|x; \phi)}[\log p(x|\text{Attention}(z, M, M); \theta_{\text{dec}})], \\
\text{Attention}(Q, K, V) &= [\text{softmax} (\frac{QW_{Q}(KW_{K})^{T}}{\sqrt{h}})VW_{V}]W_{O},
\label{eqa_vae_rec}
\end{aligned}
\end{equation}
where $\theta_{\text{dec}}$ denotes the waveform decoder, which covers not only the original waveform decoder but also the model parameters related to the memory mechanism, including the memory bank $M$ and the attention parameters $W_{Q}$, $W_{K}$, $W_{V}$, and $W_{O}$, where $M \in \mathbb{R}^{L\times h}$ and $W_{*} \in \mathbb{R}^{h\times h}$, $L$ is the size of the memory bank and $h$ is the hidden dimension.

\subsection{Training and Inference Pipeline}
\label{sec_train_infer}

Besides the waveform reconstruction loss and bidirectional prior/posterior loss, we additionally conduct a fully end-to-end optimization to take the whole inference procedure in training for better voice quality. The loss function is formulated as follows. 
\begin{equation}
\begin{aligned}
\mathcal{L}_{\text{e2e}}(\theta_{\text{pri}}, \theta_{\text{bpp}}, \theta_{\text{dec}}) & = -\mathbb{E}_{z' \sim p(z'|y; \theta_{\text{pri}})}[\log p(x|\text{Attention}(f(z'; \theta_{\text{bpp}}), M, M); \theta_{\text{dec}})].
\end{aligned}
\label{eqa_fully_e2e}
\end{equation}

Based on Equation~\ref{eqa_kl_backward},~\ref{eqa_kl_forward},~\ref{eqa_vae_rec}, and ~\ref{eqa_fully_e2e}, the total loss function is
\begin{equation}
\begin{aligned}
\mathcal{L} = \mathcal{L}_{\text{bwd}}(\phi,\theta_{\text{pri}}, \theta_{\text{bpp}}) + \mathcal{L}_{\text{fwd}}(\phi,\theta_{\text{pri}}, \theta_{\text{bpp}}) + \mathcal{L}_{\text{rec}}(\phi, \theta_{\text{dec}}) + \mathcal{L}_{\text{e2e}}(\theta_{\text{pri}}, \theta_{\text{bpp}}, \theta_{\text{dec}}),
\end{aligned}
\label{eqa_total_loss}
\end{equation}
where $\theta_{\text{pri}} = [\theta_{\text{pho}}, \theta_{\text{dur}}]$.

\begin{wrapfigure}{r}{4cm}
  \centering
\includegraphics[page=4,width=0.3\columnwidth,trim=6.3cm 1.9cm 6.7cm 2.4cm,clip=true]{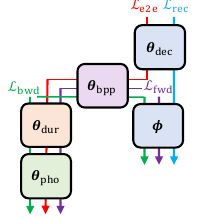}
\caption{Gradient flows.}
\label{fig_gradient_flow}
\vspace{-2mm}
\end{wrapfigure}
Note that there are some special explanations of the above loss functions: 1) Since the frame-level prior distribution $p(z'|y; \theta_{\text{pri}})$ cannot well align with the ground-truth speech frames due to the intrinsically inaccurate duration prediction in durator, we leverage a soft dynamic time warping (DTW) version of KL loss for $\mathcal{L}_{\text{bwd}}$ and $\mathcal{L}_{\text{fwd}}$. See Appendix~\ref{appendix_soft_dtw} for the detailed formulation of the soft-DTW loss. 2) We write the waveform loss in  $\mathcal{L}_{\text{rec}}$ and $\mathcal{L}_{\text{e2e}}$ as negative log-likelihood loss for simplicity. Actually following~\cite{kong2020hifi}, $\mathcal{L}_{\text{rec}}$ consists of GAN loss, feature mapping loss and mel-spectrogram loss, while $\mathcal{L}_{\text{e2e}}$ consists of only GAN loss. We do not use soft-DTW in $\mathcal{L}_{\text{e2e}}$ since we found GAN loss can still perform well with mismatched lengths. See Appendix~\ref{appendix_wav_dec_loss} for the details of the waveform loss.

There are several different gradient flows in training the model, as shown in Figure~\ref{fig_gradient_flow}: 1) $\mathcal{L}_{\text{rec}} \rightarrow  \theta_{\text{dec}} \rightarrow \phi$; 2) $\mathcal{L}_{\text{bwd}} \rightarrow \theta_{\text{dur}} \rightarrow \theta_{\text{pho}}$; 3) $\mathcal{L}_{\text{bwd}} \rightarrow \theta_{\text{bpp}} \rightarrow \phi$; 4) $\mathcal{L}_{\text{fwd}} \rightarrow \theta_{\text{bpp}} \rightarrow \theta_{\text{dur}} \rightarrow \theta_{\text{pho}}$; 5)  $\mathcal{L}_{\text{fwd}} \rightarrow \phi$; 6) $\mathcal{L}_{\text{e2e}} \rightarrow  \theta_{\text{dec}} \rightarrow \theta_{\text{bpp}} \rightarrow \theta_{\text{dur}} \rightarrow \theta_{\text{pho}}$. After training, we discard the posterior encoder $\phi$ and only use $\theta_{\text{pho}}, \theta_{\text{bpp}}, \theta_{\text{dur}}$ and $\theta_{\text{dec}}$ for inference. The training and inference pipeline is summarized in Algorithm~\ref{alg_train_infer}.

\begin{algorithm}[h]
\caption{Training and inference of \myname{}}
\label{alg}
\begin{algorithmic}[1]
\State \textbf{Training:}
\State ~~~~ Pre-train the phoneme encoder $\theta_{\text{pho}}$. 
\State ~~~~ Train the whole model $[\phi, \theta_{\text{pho}}, \theta_{\text{dur}}, \theta_{\text{bpp}}, \theta_{\text{dec}}]$ using loss $\mathcal{L}$ defined in Equation~\ref{eqa_total_loss}.
\State \textbf{Inference:} 
\State ~~~~ Sample prior $z' \sim p(z'|y; \theta_{\text{pho}}, \theta_{\text{dur}})$.
\State ~~~~ Get enhanced prior $z = f(z'; \theta_\text{bpp})$.
\State ~~~~ Generate waveform sample $x \sim p(x|\text{Attention}(z, M, M); \theta_\text{dec})$. 
\end{algorithmic}
\label{alg_train_infer}
\end{algorithm}

\subsection{Advantages of \myname{}}
\label{sec_why_can_close}

We explain how the designs in our \myname{} system can close the quality gap to recordings.

\begin{itemize}[leftmargin=*]
\item \textit{Reduce training-inference mismatch}. We directly generate waveform from text and leverage a differentiable durator to ensure a fully end-to-end optimization, which can reduce the training-inference mismatch in the cascaded acoustic model/vocoder~\citep{kim2020glow,ren2021fastspeech,popov2021grad,elias2021parallel} and explicit duration prediction~\citep{kim2021conditional,kim2020glow,ren2021fastspeech}. Note that although VAE and flow can have training-inference mismatch inherently (waveform is reconstructed from the posterior in training while predicted from the prior in inference for VAE, and flow is trained in backward direction and infered in forward direction), we design the backward/forward loss in Equation~\ref{eqa_kl_backward} and~\ref{eqa_kl_forward} and the end-to-end loss in Equation~\ref{eqa_fully_e2e} to alleviate this problem.  

\item \textit{Alleviate one-to-many mapping problem}. Compared to previous methods using reference encoder~\citep{wang2018style,chen2021adaspeech,wu2022adaspeech,liu2021delightfultts} or pitch/energy extraction~\citep{ren2021fastspeech} for variation information modeling, our posterior encoder $\phi$ in VAE acts like a reference encoder that can extract all the necessary variance information in posterior distribution $q(z|x; \phi)$. We do not predict pitch explicitly since it can be learned implicitly in the posterior encoder and the memory bank of VAE. To ensure the prior and posterior can match with each other, on the one hand, we simplify the posterior with memory VAE and backward mapping in the bidirectional prior/posterior module, and on the other hand, we enhance the prior with phoneme pre-training, differentiable durator, and forward mapping in the bidirectional prior/posterior module. Thus, we can alleviate the one-to-mapping problem to a large extent. 

\item \textit{Increase representation capacity}. We leverage large-scale phoneme pre-training to extract better representation from the phoneme sequence, and leverage the advanced generative models (flow, VAE, GAN) to capture the speech data distributions better, which can enhance the representation capacity of the TTS models for better voice quality. 

\end{itemize}

We further list the difference between our \myname{} and previous TTS systems as follows: 1) Compared to previous autoregressive TTS models such as Tacotron 1/2~\citep{wang2017tacotron,shen2018natural}, WaveNet~\citep{oord2016wavenet}, TransformerTTS~\citep{li2019neural}, and Wave-Tacotron~\cite{weiss2021wave}, our \myname{} is non-autoregressive in nature with a fast inference speed. 2) Compared to the previous systems with cascaded acoustic model and vocoder, such as Tacotron 1/2~\citep{wang2017tacotron,shen2018natural}, FastSpeech 1/2~\citep{ren2019fastspeech,ren2021fastspeech}, ParallelTacotron 2~\citep{elias2021parallel}, Glow-TTS~\citep{kim2020glow}, and Grad-TTS~\citep{popov2021grad}, we are fully end-to-end with no cascaded errors. 3) Compared to previous systems with various reference encoders and pitch/duration prediction, such as FastSpeech 2~\citep{ren2021fastspeech}, AdaSpeech~\citep{chen2021adaspeech}, and DelightfulTTS~\citep{liu2021delightfultts}, we unify all the variance information with a posterior encoder and model the duration in a fully differentiable way. 4) Compared to previous fully end-to-end TTS systems such as EATS~\citep{donahue2020end}, FastSpeech 2s~\citep{ren2021fastspeech}, and VITS~\citep{kim2021conditional}, we bridge the quality gap to recordings with advanced model designs to closely match the prior and posterior in the VAE framework.

\section{Experiments and Results}

\subsection{Experimental Settings}
\label{sec_exp_setting}

\paragraph{Datasets}
We evaluate our proposed \myname{} on the LJSpeech dataset~\citep{ljspeech17}, which is widely used for benchmarking TTS. LJSpeech is a single speaker English corpus and consists of $13,100$ audios and text transcripts, with a total length of nearly 24 hours at a sampling rate of $22.05$kHz. We randomly split the dataset into training set with $12,500$ samples, validation set with $100$ samples, and test set with $500$ samples. For phoneme pre-training on phoneme encoder, we collect a large-scale text corpus with $200$ million sentences from the news-crawl dataset~\citep{newscrawl22}. Note that we do not use any extra paired text and speech data except for LJSpeech dataset. We conduct several preprocessings on the speech and text sequences: 1) We convert the text/character sequence into phoneme sequence~\cite{sun2019token} using a grapheme-to-phoneme tool~\citep{Bernard2021}. 2) We use linear-spectrograms as the input of the posterior encoder~\citep{kim2021conditional}, instead of original waveform sequence for simplicity. The linear-spectrograms are obtained by short-time Fourier transform (STFT) with FFT size, window size, and hop size of 1024, 1024, and 256, respectively. 3) For the mel-spectrogram loss on the waveform decoder, we obtain the mel-spectrograms by applying $80$-dimension mel-filterbanks on the linear-spectrograms of the speech waveform. 

\paragraph{Model Configurations} 
Our phoneme encoder is a stack of $6$ Feed-Forward Transformer~(FFT) blocks~\cite{ren2019fastspeech}, where each block consists of a multi-head attention layer and a 1D convolution feed-forward layer, with hidden size of $192$. In the differentiable durator, the duration predictor consists of $3$-layer convolution. We use 4 consecutive affine coupling layers~\cite{dinh2016density} in our bidirectional prior/posterior module following~\citep{kim2021conditional}. We discard the scaling operation in the affine transform to stabilize the bidirectional training. The shifting in the affine transform is estimated by a $4$-layer WaveNet~\cite{oord2016wavenet} with a dilation rate of $1$. The posterior encoder is based on a $16$-layer WaveNet with a kernel size of $5$ and a dilation rate of $1$. The waveform decoder consists of $4$ residual convolution blocks following~\citep{kong2020hifi}, where each block has $3$ layers of 1D convolution. We perform transpose convolution for upsampling at every convolution block at a rate of $[8,8,2,2]$. The hyperparameters of \myname{} are listed in Appendix~\ref{appendix_hyperpara}.

\paragraph{Training Details}
\label{sec:training-details}
We train our proposed system on $8$ NVIDIA V100 GPUs with 32G memory, with a dynamic batch size of $8,000$ speech frames (under hop size of $256$) per GPU, and a total $15$k training epochs. We use AdamW optimizer~\citep{loshchilov2018decoupled} with $\beta_1$ = $0.8$, $\beta_2$ = $0.99$. The initial learning rate is $2 \times 10^{-4}$, with a learning rate decay factor $\gamma=0.999875$ in each epoch, i.e., the learning rate is multiplied by $\gamma$ in every epoch. We find it is helpful to stabilize the training of our system and achieve better results through a warmup stage with $1$k epochs at the beginning of the training, and a tuning stage with $2$k epochs at the end of the training. More details about these training stages can be found in Appendix~\ref{appendix_training_details}.

\subsection{Comparison with Human Recordings}
We first compare the speech generated by \myname{} with human recordings in terms of both MOS and CMOS evaluation. As described in Section~\ref{sec_def_judge}, we use 50 test utterances, each with 20 judges for evaluation. As shown in Table~\ref{tab_mos_recording} and ~\ref{tab_cmos_recording}, our system achieves similar quality scores with human recordings in both MOS and CMOS. Importantly, our system achieves $-0.01$ CMOS compared to recordings, with a Wilcoxon p-value~\citep{wilcoxon1992individual} $p \gg 0.05$, which demonstrates the speech generated by our system has no statistically significant difference from human recordings\footnote{Audio samples can be found in \url{https://speechresearch.github.io/naturalspeech/}}~\footnote{Note that some human recordings in LJSpeech dataset may contain strange rhythm ups and downs that affect the rating score. To ensure the human recordings used for evaluation are of good quality, we let judges to exclude the recordings with strange rhythms from evaluation. Otherwise, our \myname{} will achieve better CMOS than human recordings. In a CMOS test without excluding bad recordings, \myname{} achieves $+0.09$ CMOS better than recordings.}. Thus, our \myname{} achieves human-level quality according to the definition and judgement in Section~\ref{sec_def_judge}.

\begin{table}[h!]
\caption{MOS comparison between \myname{} and human recordings. Wilcoxon rank sum test is used to measure the p-value in MOS evaluation.}
\centering
\begin{tabular}{ccc}
	\toprule
	Human Recordings &\myname{} & Wilcoxon p-value \\
	\midrule
	$4.58\pm0.13$ & $4.56\pm0.13$ & $0.7145$ \\
	\bottomrule
\end{tabular}
\label{tab_mos_recording}
\end{table}

\begin{table}[h!]
\caption{CMOS comparison between \myname{} and human recordings. Wilcoxon signed rank test is used to measure the p-value in CMOS evaluation.}
\centering
\begin{tabular}{ccc}
	\toprule
	Human Recordings &\myname{} & Wilcoxon p-value \\
	\midrule
	0 & $-0.01$ & $0.6902$ \\
	\bottomrule
\end{tabular}
\label{tab_cmos_recording}
\end{table}

\subsection{Comparison with Previous TTS Systems}
We compare our \myname{} with previous TTS systems, including: 1) FastSpeech 2~\citep{ren2021fastspeech} + HiFiGAN~\citep{kong2020hifi}, 2) Glow-TTS~\citep{kim2020glow} + HiFiGAN~\citep{kong2020hifi}, 3) Grad-TTS~\citep{popov2021grad} + HiFiGAN~\citep{kong2020hifi}, and 4) VITS~\citep{kim2021conditional}. We re-produce the results of all these systems by our own, which can match or even beat the quality in their original papers (note that the HiFiGAN vocoder is fine-tuned on the predicted mel-spectrograms for better synthesis quality). Both the MOS and CMOS results are shown in Table~\ref{tab_mos_previous_system}.  It can be seen that our \myname{} achieves better voice quality than these systems in terms of both MOS and CMOS.

\begin{table}[h!]
\caption{MOS and CMOS comparisons between \myname{} and previous TTS systems.}
\centering
\begin{tabular}{ l | l | c }
    \toprule
	    System & MOS & CMOS \\
	    \midrule
	    FastSpeech 2~\citep{ren2021fastspeech} + HiFiGAN~\citep{kong2020hifi} & $4.32\pm0.15$ & $-0.33$ \\
	    Glow-TTS~\citep{kim2020glow} + HiFiGAN~\citep{kong2020hifi} & $4.34\pm0.13$ & $-0.26$ \\
	    Grad-TTS~\citep{popov2021grad} + HiFiGAN~\citep{kong2020hifi}  & $4.37\pm0.13$  & $-0.24$ \\
	    VITS~\citep{kim2021conditional} &  $4.43\pm0.13$ & $-0.20$ \\
	    \midrule
	    \myname{} & $4.56\pm0.13$ & $0$ \\
		\bottomrule
	\end{tabular}
	\label{tab_mos_previous_system}
\end{table}

\subsection{Ablation Studies and Method Analyses}
\paragraph{Ablation Studies} We further conduct ablation studies to verify the effectiveness of each module in our system, as shown in Table~\ref{tab_cmos_ablation}. We describe the ablation studies as follows: 1) By removing phoneme pre-training, we do not initialize the phoneme encoder from pre-trained weights but just random initialization, which brings $-0.09$ CMOS drop, demonstrating the effectiveness of phoneme pre-training. 2) By removing differentiable durator, we do not use learnable upsampling layer and end-to-end duration optimization, but just use duration predictor for hard expansion. In this way, we use monotonic alignment search~\citep{kim2020glow} to provide the duration label to train the duration predictor through the whole training process. Removing differentiable durator causes $-0.12$ CMOS drop, demonstrates the importance of end-to-end optimization in duration modeling. 3) By removing bidirectional prior/posterior module, we only use $\mathcal{L}_{\text{bwd}}$ in training and do not use $\mathcal{L}_{\text{fwd}}$. It brings $-0.09$ CMOS drop, showing the gain by leveraging bidirectional training to bridge the gap between posterior and prior. 4) By removing memory mechanism in VAE, we use original VAE for waveform reconstruction, which causes $-0.06$ CMOS drop, showing the effectiveness of memory in VAE to simplify the posterior.

\begin{table}[h!]
	\caption{Ablation studies on each design in \myname{} .}
	\centering
	\begin{tabular}{ l c}
		\toprule
	    Setting & CMOS \\
	    \midrule
		\myname{}  & $0$ \\
		$-$ Phoneme Pre-training & $-0.09$ \\
		$-$ Differentiable Durator & $-0.12$\\
		$-$ Bidirectional Prior/Posterior & $-0.09$ \\
		$-$ Memory in VAE & $-0.06$ \\
		\bottomrule
	\end{tabular}
	\label{tab_cmos_ablation}
\end{table}

\paragraph{Inference Latency}
We compare the inference speed of our \myname{} with previous TTS systems. We measure the latency by using an NVIDIA V100 GPU with a batch size of $1$ sentence and averaging the latency over the sentences in the test set. The results are shown in Table~\ref{tab_exp_latency}. The model components $\theta_{\rm pho}$, $\theta_{\rm dur}$, $\theta_{\rm bpp}$, and $\theta_{\rm dec}$ in \myname{} are used in inference, with $28.7$M model parameters. Our \myname{} achieves faster or comparable inference speed when compared with the previous systems, and achieves better voice quality.

\begin{table}[h!]
	\caption{Inference speed comparison. RTF (real-time factor) means the time (in seconds) to synthesize a $1$-second waveform.  Grad-TTS (1000) and  Grad-TTS (10) mean using 1000 and 10 steps in inference respectively.}
	\centering
	\begin{tabular}{ l c}
		\toprule
	    System & RTF \\
	    \midrule
	    FastSpeech 2~\citep{ren2021fastspeech} + HiFiGAN~\citep{kong2020hifi}  & $0.011$ \\
	    Glow-TTS~\citep{kim2020glow} + HiFiGAN~\citep{kong2020hifi} & $0.021$\\
	    Grad-TTS~\citep{popov2021grad} (1000) + HiFiGAN~\citep{kong2020hifi} & $4.120$\\
	    Grad-TTS~\citep{popov2021grad} (10) + HiFiGAN~\citep{kong2020hifi} & $0.082$\\
	    VITS~\citep{kim2021conditional} & $0.014$\\
	    \midrule
		\myname{}  & $0.013$ \\
		\bottomrule
	\end{tabular}
	\label{tab_exp_latency}
\end{table}

\section{Conclusions and Discussions}
In this paper, we conduct a systematic study on the problems related to human-level quality in TTS. We first give a formal definition of human-level quality and describe the guidelines to judge it, and further build a TTS system called \myname{} to achieve human-level quality. Specifically, after analyzing the quality gap on several competitive TTS systems, we develop a fully end-to-end text to waveform generation system, with several designs to close the gap to human recordings, including phoneme pre-training, differentiable durator, bidirectional prior/posterior module, and memory mechanism in VAE. Evaluations on the popular LJSpeech dataset demonstrate that our \myname{} achieves human-level quality with CMOS evaluations, with no statistically significant difference from human recordings for the first time on this dataset. 

Note that by claiming our \myname{} system achieves human-level quality on LJSpeech dataset, we do not mean that we can surpass or replace human, but the quality of \myname{} is statistically indistinguishable from human recordings on this dataset. Meanwhile, although our evaluations are conducted on LJSpeech dataset, we believe the technologies in \myname{} can be applied to other languages, speakers, and styles to improve the general synthesis quality. We will further try to achieve human-level quality in more challenging datasets or scenarios, such as expressive voices, longform audiobook voices, and singing voices that have more dynamic, diverse, and contextual prosody in our future work.

\bibliography{main}

\begin{thebibliography}{10}

\bibitem{tan2021survey}
Xu~Tan, Tao Qin, Frank Soong, and Tie-Yan Liu.
\newblock A survey on neural speech synthesis.
\newblock {\em arXiv preprint arXiv:2106.15561}, 2021.

\bibitem{oord2016wavenet}
Aaron van~den Oord, Sander Dieleman, Heiga Zen, Karen Simonyan, Oriol Vinyals,
  Alex Graves, Nal Kalchbrenner, Andrew Senior, and Koray Kavukcuoglu.
\newblock Wavenet: A generative model for raw audio.
\newblock {\em arXiv preprint arXiv:1609.03499}, 2016.

\bibitem{shen2018natural}
Jonathan Shen, Ruoming Pang, Ron~J Weiss, Mike Schuster, Navdeep Jaitly,
  Zongheng Yang, Zhifeng Chen, Yu~Zhang, Yuxuan Wang, Rj~Skerrv-Ryan, et~al.
\newblock Natural tts synthesis by conditioning wavenet on mel spectrogram
  predictions.
\newblock In {\em 2018 IEEE International Conference on Acoustics, Speech and
  Signal Processing (ICASSP)}, pages 4779--4783. IEEE, 2018.

\bibitem{wang2017tacotron}
Yuxuan Wang, RJ~Skerry-Ryan, Daisy Stanton, Yonghui Wu, Ron~J Weiss, Navdeep
  Jaitly, Zongheng Yang, Ying Xiao, Zhifeng Chen, Samy Bengio, et~al.
\newblock Tacotron: Towards end-to-end speech synthesis.
\newblock {\em Proc. Interspeech 2017}, pages 4006--4010, 2017.

\bibitem{arik2017deep}
Sercan~{\"O} Ar{\i}k, Mike Chrzanowski, Adam Coates, Gregory Diamos, Andrew
  Gibiansky, Yongguo Kang, Xian Li, John Miller, Andrew Ng, Jonathan Raiman,
  et~al.
\newblock Deep voice: Real-time neural text-to-speech.
\newblock In {\em International Conference on Machine Learning}, pages
  195--204. PMLR, 2017.

\bibitem{gibiansky2017deep}
Andrew Gibiansky, Sercan~{\"O}mer Arik, Gregory~Frederick Diamos, John Miller,
  Kainan Peng, Wei Ping, Jonathan Raiman, and Yanqi Zhou.
\newblock Deep voice 2: Multi-speaker neural text-to-speech.
\newblock In {\em NIPS}, 2017.

\bibitem{ping2018deep}
Wei Ping, Kainan Peng, Andrew Gibiansky, Sercan~O Arik, Ajay Kannan, Sharan
  Narang, Jonathan Raiman, and John Miller.
\newblock Deep voice 3: 2000-speaker neural text-to-speech.
\newblock {\em Proc. ICLR}, pages 214--217, 2018.

\bibitem{tachibana2018efficiently}
Hideyuki Tachibana, Katsuya Uenoyama, and Shunsuke Aihara.
\newblock Efficiently trainable text-to-speech system based on deep
  convolutional networks with guided attention.
\newblock In {\em 2018 IEEE International Conference on Acoustics, Speech and
  Signal Processing (ICASSP)}, pages 4784--4788. IEEE, 2018.

\bibitem{li2019neural}
Naihan Li, Shujie Liu, Yanqing Liu, Sheng Zhao, and Ming Liu.
\newblock Neural speech synthesis with transformer network.
\newblock In {\em Proceedings of the AAAI Conference on Artificial
  Intelligence}, volume~33, pages 6706--6713, 2019.

\bibitem{ren2019fastspeech}
Yi~Ren, Yangjun Ruan, Xu~Tan, Tao Qin, Sheng Zhao, Zhou Zhao, and Tie-Yan Liu.
\newblock Fastspeech: Fast, robust and controllable text to speech.
\newblock In {\em NeurIPS}, 2019.

\bibitem{liu2021delightfultts}
Yanqing Liu, Zhihang Xu, Gang Wang, Kuan Chen, Bohan Li, Xu~Tan, Jinzhu Li, Lei
  He, and Sheng Zhao.
\newblock Delightfultts: The microsoft speech synthesis system for blizzard
  challenge 2021.
\newblock {\em arXiv preprint arXiv:2110.12612}, 2021.

\bibitem{prenger2019waveglow}
Ryan Prenger, Rafael Valle, and Bryan Catanzaro.
\newblock Waveglow: A flow-based generative network for speech synthesis.
\newblock In {\em ICASSP 2019-2019 IEEE International Conference on Acoustics,
  Speech and Signal Processing (ICASSP)}, pages 3617--3621. IEEE, 2019.

\bibitem{kim2020glow}
Jaehyeon Kim, Sungwon Kim, Jungil Kong, and Sungroh Yoon.
\newblock Glow-tts: A generative flow for text-to-speech via monotonic
  alignment search.
\newblock {\em Advances in Neural Information Processing Systems}, 33, 2020.

\bibitem{popov2021grad}
Vadim Popov, Ivan Vovk, Vladimir Gogoryan, Tasnima Sadekova, and Mikhail
  Kudinov.
\newblock Grad-tts: A diffusion probabilistic model for text-to-speech.
\newblock In {\em International Conference on Machine Learning}, pages
  8599--8608. PMLR, 2021.

\bibitem{kim2021conditional}
Jaehyeon Kim, Jungil Kong, and Juhee Son.
\newblock Conditional variational autoencoder with adversarial learning for
  end-to-end text-to-speech.
\newblock In {\em International Conference on Machine Learning}, pages
  5530--5540. PMLR, 2021.

\bibitem{kalchbrenner2018efficient}
Nal Kalchbrenner, Erich Elsen, Karen Simonyan, Seb Noury, Norman Casagrande,
  Edward Lockhart, Florian Stimberg, Aaron Oord, Sander Dieleman, and Koray
  Kavukcuoglu.
\newblock Efficient neural audio synthesis.
\newblock In {\em International Conference on Machine Learning}, pages
  2410--2419. PMLR, 2018.

\bibitem{kong2020hifi}
Jungil Kong, Jaehyeon Kim, and Jaekyoung Bae.
\newblock Hifi-gan: Generative adversarial networks for efficient and high
  fidelity speech synthesis.
\newblock {\em Advances in Neural Information Processing Systems}, 33, 2020.

\bibitem{ren2021fastspeech}
Yi~Ren, Chenxu Hu, Xu~Tan, Tao Qin, Sheng Zhao, Zhou Zhao, and Tie-Yan Liu.
\newblock Fastspeech 2: Fast and high-quality end-to-end text to speech.
\newblock In {\em International Conference on Learning Representations}, 2021.

\bibitem{donahue2020end}
Jeff Donahue, Sander Dieleman, Miko{\l}aj Bi{\'n}kowski, Erich Elsen, and Karen
  Simonyan.
\newblock End-to-end adversarial text-to-speech.
\newblock In {\em ICLR}, 2021.

\bibitem{van2017neural}
Aaron van~den Oord, Oriol Vinyals, and Koray Kavukcuoglu.
\newblock Neural discrete representation learning.
\newblock In {\em Proceedings of the 31st International Conference on Neural
  Information Processing Systems}, pages 6309--6318, 2017.

\bibitem{ramesh2021zero}
Aditya Ramesh, Mikhail Pavlov, Gabriel Goh, Scott Gray, Chelsea Voss, Alec
  Radford, Mark Chen, and Ilya Sutskever.
\newblock Zero-shot text-to-image generation.
\newblock {\em arXiv preprint arXiv:2102.12092}, 2021.

\bibitem{kingma2013auto}
Diederik~P Kingma and Max Welling.
\newblock Auto-encoding variational bayes.
\newblock {\em arXiv preprint arXiv:1312.6114}, 2013.

\bibitem{dinh2014nice}
Laurent Dinh, David Krueger, and Yoshua Bengio.
\newblock Nice: Non-linear independent components estimation.
\newblock {\em arXiv preprint arXiv:1410.8516}, 2014.

\bibitem{kingma2016improved}
Durk~P Kingma, Tim Salimans, Rafal Jozefowicz, Xi~Chen, Ilya Sutskever, and Max
  Welling.
\newblock Improved variational inference with inverse autoregressive flow.
\newblock {\em Advances in Neural Information Processing Systems},
  29:4743--4751, 2016.

\bibitem{kingma2018glow}
Diederik~P Kingma and Prafulla Dhariwal.
\newblock Glow: generative flow with invertible 1$\times$ 1 convolutions.
\newblock In {\em Proceedings of the 32nd International Conference on Neural
  Information Processing Systems}, pages 10236--10245, 2018.

\bibitem{ljspeech17}
Keith Ito.
\newblock The lj speech dataset.
\newblock \url{https://keithito.com/LJ-Speech-Dataset/}, 2017.

\bibitem{rix2001perceptual}
Antony~W Rix, John~G Beerends, Michael~P Hollier, and Andries~P Hekstra.
\newblock Perceptual evaluation of speech quality (pesq)-a new method for
  speech quality assessment of telephone networks and codecs.
\newblock In {\em 2001 IEEE international conference on acoustics, speech, and
  signal processing. Proceedings (Cat. No. 01CH37221)}, volume~2, pages
  749--752. IEEE, 2001.

\bibitem{taal2011algorithm}
Cees~H Taal, Richard~C Hendriks, Richard Heusdens, and Jesper Jensen.
\newblock An algorithm for intelligibility prediction of time--frequency
  weighted noisy speech.
\newblock {\em IEEE Transactions on Audio, Speech, and Language Processing},
  19(7):2125--2136, 2011.

\bibitem{le2019sdr}
Jonathan Le~Roux, Scott Wisdom, Hakan Erdogan, and John~R Hershey.
\newblock Sdr--half-baked or well done?
\newblock In {\em ICASSP 2019-2019 IEEE International Conference on Acoustics,
  Speech and Signal Processing (ICASSP)}, pages 626--630. IEEE, 2019.

\bibitem{wilcoxon1992individual}
Frank Wilcoxon.
\newblock Individual comparisons by ranking methods.
\newblock In {\em Breakthroughs in statistics}, pages 196--202. Springer, 1992.

\bibitem{ding2021cogview}
Ming Ding, Zhuoyi Yang, Wenyi Hong, Wendi Zheng, Chang Zhou, Da~Yin, Junyang
  Lin, Xu~Zou, Zhou Shao, Hongxia Yang, et~al.
\newblock Cogview: Mastering text-to-image generation via transformers.
\newblock {\em arXiv preprint arXiv:2105.13290}, 2021.

\bibitem{wu2021nuwa}
Chenfei Wu, Jian Liang, Lei Ji, Fan Yang, Yuejian Fang, Daxin Jiang, and Nan
  Duan.
\newblock N\"uwa: Visual synthesis pre-training for neural visual world
  creation, 2021.

\bibitem{yan2021videogpt}
Wilson Yan, Yunzhi Zhang, Pieter Abbeel, and Aravind Srinivas.
\newblock Videogpt: Video generation using vq-vae and transformers.
\newblock {\em arXiv preprint arXiv:2104.10157}, 2021.

\bibitem{rakhimov2020latent}
Ruslan Rakhimov, Denis Volkhonskiy, Alexey Artemov, Denis Zorin, and Evgeny
  Burnaev.
\newblock Latent video transformer.
\newblock {\em arXiv preprint arXiv:2006.10704}, 2020.

\bibitem{razavi2019generating}
Ali Razavi, Aaron van~den Oord, and Oriol Vinyals.
\newblock Generating diverse high-fidelity images with vq-vae-2.
\newblock In {\em Advances in neural information processing systems}, pages
  14866--14876, 2019.

\bibitem{esser2021taming}
Patrick Esser, Robin Rombach, and Bjorn Ommer.
\newblock Taming transformers for high-resolution image synthesis.
\newblock In {\em Proceedings of the IEEE/CVF Conference on Computer Vision and
  Pattern Recognition}, pages 12873--12883, 2021.

\bibitem{vaswani2017attention}
Ashish Vaswani, Noam Shazeer, Niki Parmar, Jakob Uszkoreit, Llion Jones,
  Aidan~N Gomez, {\L}ukasz Kaiser, and Illia Polosukhin.
\newblock Attention is all you need.
\newblock In {\em Advances in Neural Information Processing Systems}, pages
  5998--6008, 2017.

\bibitem{xiao2020improving}
Yujia Xiao, Lei He, Huaiping Ming, and Frank~K Soong.
\newblock Improving prosody with linguistic and bert derived features in
  multi-speaker based mandarin chinese neural tts.
\newblock In {\em ICASSP 2020-2020 IEEE International Conference on Acoustics,
  Speech and Signal Processing (ICASSP)}, pages 6704--6708. IEEE, 2020.

\bibitem{jia2021png}
Ye~Jia, Heiga Zen, Jonathan Shen, Yu~Zhang, and Yonghui Wu.
\newblock Png bert: Augmented bert on phonemes and graphemes for neural tts.
\newblock {\em Proc. Interspeech 2021}, pages 151--155, 2021.

\bibitem{zhang2022mixed}
Guangyan Zhang, Kaitao Song, Xu~Tan, Daxin Tan, Yuzi Yan, Yanqing Liu, Gang
  Wang, Wei Zhou, Tao Qin, Tan Lee, et~al.
\newblock Mixed-phoneme bert: Improving bert with mixed phoneme and sup-phoneme
  representations for text to speech.
\newblock {\em arXiv preprint arXiv:2203.17190}, 2022.

\bibitem{devlin2018bert}
Jacob Devlin, Ming-Wei Chang, Kenton Lee, and Kristina Toutanova.
\newblock Bert: Pre-training of deep bidirectional transformers for language
  understanding.
\newblock {\em arXiv preprint arXiv:1810.04805}, 2018.

\bibitem{elias2021parallel}
Isaac Elias, Heiga Zen, Jonathan Shen, Yu~Zhang, Jia Ye, RJ~Ryan, and Yonghui
  Wu.
\newblock Parallel tacotron 2: A non-autoregressive neural tts model with
  differentiable duration modeling.
\newblock {\em arXiv preprint arXiv:2103.14574}, 2021.

\bibitem{rezende2015variational}
Danilo Rezende and Shakir Mohamed.
\newblock Variational inference with normalizing flows.
\newblock In {\em International Conference on Machine Learning}, pages
  1530--1538. PMLR, 2015.

\bibitem{wang2018style}
Yuxuan Wang, Daisy Stanton, Yu~Zhang, RJ-Skerry Ryan, Eric Battenberg, Joel
  Shor, Ying Xiao, Ye~Jia, Fei Ren, and Rif~A Saurous.
\newblock Style tokens: Unsupervised style modeling, control and transfer in
  end-to-end speech synthesis.
\newblock In {\em International Conference on Machine Learning}, pages
  5180--5189. PMLR, 2018.

\bibitem{chen2021adaspeech}
Mingjian Chen, Xu~Tan, Bohan Li, Yanqing Liu, Tao Qin, sheng zhao, and Tie-Yan
  Liu.
\newblock Adaspeech: Adaptive text to speech for custom voice.
\newblock In {\em International Conference on Learning Representations}, 2021.

\bibitem{wu2022adaspeech}
Yihan Wu, Xu~Tan, Bohan Li, Lei He, Sheng Zhao, Ruihua Song, Tao Qin, and
  Tie-Yan Liu.
\newblock Adaspeech 4: Adaptive text to speech in zero-shot scenarios.
\newblock {\em arXiv preprint arXiv:2204.00436}, 2022.

\bibitem{weiss2021wave}
Ron~J Weiss, RJ~Skerry-Ryan, Eric Battenberg, Soroosh Mariooryad, and
  Diederik~P Kingma.
\newblock Wave-tacotron: Spectrogram-free end-to-end text-to-speech synthesis.
\newblock In {\em ICASSP 2021-2021 IEEE International Conference on Acoustics,
  Speech and Signal Processing (ICASSP)}, pages 5679--5683. IEEE, 2021.

\bibitem{newscrawl22}
Machine Translation~Group at~UEDIN.
\newblock The news-crawl dataset.
\newblock \url{https://data.statmt.org/news-crawl/en/}, 2022.

\bibitem{sun2019token}
Hao Sun, Xu~Tan, Jun-Wei Gan, Hongzhi Liu, Sheng Zhao, Tao Qin, and Tie-Yan
  Liu.
\newblock Token-level ensemble distillation for grapheme-to-phoneme conversion.
\newblock In {\em INTERSPEECH}, 2019.

\bibitem{Bernard2021}
Mathieu Bernard and Hadrien Titeux.
\newblock Phonemizer: Text to phones transcription for multiple languages in
  python.
\newblock {\em Journal of Open Source Software}, 6(68):3958, 2021.

\bibitem{dinh2016density}
Laurent Dinh, Jascha Sohl-Dickstein, and Samy Bengio.
\newblock Density estimation using real nvp.
\newblock {\em arXiv preprint arXiv:1605.08803}, 2016.

\bibitem{loshchilov2018decoupled}
Ilya Loshchilov and Frank Hutter.
\newblock Decoupled weight decay regularization.
\newblock In {\em International Conference on Learning Representations}, 2018.

\bibitem{ramachandran2017searching}
Prajit Ramachandran, Barret Zoph, and Quoc~V Le.
\newblock Searching for activation functions.
\newblock {\em arXiv:1710.05941}, 2017.

\bibitem{mao2017least}
Xudong Mao, Qing Li, Haoran Xie, Raymond~YK Lau, Zhen Wang, and Stephen
  Paul~Smolley.
\newblock Least squares generative adversarial networks.
\newblock In {\em Proceedings of the IEEE international conference on computer
  vision}, pages 2794--2802, 2017.

\bibitem{sennrich2015neural}
Rico Sennrich, Barry Haddow, and Alexandra Birch.
\newblock Neural machine translation of rare words with subword units.
\newblock {\em arXiv preprint arXiv:1508.07909}, 2015.

\bibitem{mcauliffe2017montreal}
Michael McAuliffe, Michaela Socolof, Sarah Mihuc, Michael Wagner, and Morgan
  Sonderegger.
\newblock Montreal forced aligner: Trainable text-speech alignment using kaldi.
\newblock In {\em Interspeech}, volume 2017, pages 498--502, 2017.

\end{thebibliography}
\bibliographystyle{unsrt}

\appendix

\section{Study of the Quality Gap of Previous TTS System}
\label{appendix_study}
To understand where and how the quality gap to recordings comes from, we conduct a systematic study on the current TTS systems, which can help us to find the problems, and is equally important (if not more) than solving the problems. Specifically, we choose a state-of-the-art TTS system using FastSpeech 2~\cite{ren2021fastspeech} as the acoustic model and HiFiGAN~\cite{kong2020hifi} as the vocoder, which consists of four components: phoneme encoder, variance adaptor, mel-spectrogram decoder, and vocoder. We design a series of comparison experiments to measure the quality gap (in terms of CMOS) of each component to its corresponding upper bound. We conduct analyses from this order (from the closest to waveform to the farest): vocoder, mel-spectrogram decoder, variance adaptor, and phoneme encoder. 

\begin{table}[h!]
\small
	\caption{The CMOS of each component to its upper bound. Negative CMOS means this component setting is worse than its upper bound.}
	\centering
	\begin{tabular}{ l l l l}
		\toprule
	    Component & Setting & Upper Bound & CMOS   \\
		\midrule
		Vocoder  & GT Mel$\rightarrow$Vocoder & Human Recordings & $-0.04$ \\
		Mel Decoder & GT Pitch/Duration$\rightarrow$Mel Decoder & GT Mel &  $-0.15$ \\
		Variance Adaptor & Predicted Pitch/Duration & GT Pitch/Duration & $-0.14$ \\
		Phoneme Encoder & Phoneme Encoder & Phoneme Encoder + Pre-training & $-0.12$ \\
		\bottomrule
	\end{tabular}
	\vspace{0.3cm}
	\label{tab_study_on_quality_drop}
\end{table}

\begin{itemize}[leftmargin=*]

\item \textit{Vocoder.} We study the quality drop on the vocoder by comparing the two settings: 1) waveform generated by vocoder with ground-truth mel-spectrograms as input; 2) ground-truth waveform (human recordings). The CMOS is shown in Table~\ref{tab_study_on_quality_drop}. It can be seen that when taking ground-truth mel-spectrograms as input, the waveform generated by vocoder has some but not huge gap to human recordings. However, we need to pay attention to the training-inference mismatch in vocoder: in training, vocoder takes ground-truth mel-spectrograms as input, while in inference, it takes predicted mel-spectrograms as input. 

\item \textit{Mel-spectrogram Decoder.} We study the quality drop on the mel-spectrogram decoder by comparing the two settings: 1) mel-spectrograms generated by mel-spectrogram decoder with ground-truth pitch and duration as input\footnote{Ideally, we should also use ground-truth phoneme hidden sequence as input. However, ground-truth hidden sequence cannot be obtained. Thus, this comparison setting is just a approximation.}; 2) ground-truth mel-spectrograms (extracted from human recordings). We use the vocoder to convert the mel-sepctrograms in the two settings into waveform for evaluation. As shown in Table~\ref{tab_study_on_quality_drop}, the predicted mel-spectrograms have $0.15$ CMOS drop compared to the ground-truth mel-spectrograms.

\item \textit{Variance Adaptor.} We study the quality drop on the variance adaptor by comparing the predicted pitch/duration with the ground-truth pitch/duration. We need the mel-spectrogram decoder and vocoder to generate the waveform for evaluation in the two settings. As shown in Table~\ref{tab_study_on_quality_drop}, the predicted pitch/duration have $0.14$ CMOS drop compared to the ground-truth pitch/duration. 

\item \textit{Phoneme Encoder.} Since it is not straightforward to construct the upper bound of the phoneme encoder, we analyze the approximate quality drop through backward verification, by improving phoneme encoder for better voice quality. We conduct large-scale phoneme pre-training on the phoneme encoder, and fine-tune it with the FastSpeech 2 training pipeline, and achieves a $0.12$ CMOS gain, as shown in Table~\ref{tab_study_on_quality_drop}, which demonstrates the phoneme encoder has improvement space.

\end{itemize}

According to the above experimental studies, we analyze several reasons causing the quality drop in each component: 1) Training-inference mismatch. Ground-truth mel-spectrogram, pitch, and duration are used in training, while predicted values are used in inference, which causes mismatch in the input of vocoder and mel-spectrogram decoder. Fully end-to-end text to waveform optimization is helpful to eliminate this mismatch. 2) One-to-many mapping problem. Text to speech mapping is one-to-many, where a text sequence can correspond to multiple speech utterances with different variation information (e.g., pitch, duration, speed, pause, prosody, etc). Current systems usually use a variance adaptor to predict variance information (e.g., pitch, duration) to alleviate this problem, which is not enough to well handle this problem. We should rethink previous methods on variance information and come up with some thorough and elegant solutions. 3) Lack of representation capacity. Current models are not powerful enough to extract good representations from phoneme sequence and learn complicated data distribution in speech. More advanced methods such as large-scale pre-training and powerful generative models are critical to enhance the learning capacity.

\section{Differentiable Durator}
\label{appendix_durator}

To enable end-to-end duration optimization, we design a durator that can upsample a phoneme hidden sequence $\mH_{n \times h}$ into a frame-level hidden sequence $\mO_{m \times h}$ in a differentiable way, where $h, n, m$ is the hidden dimension size, phoneme sequence length and frame sequence length, respectively.
The differentiable durator consists of a duration predictor $\theta_{\rm dp}$ for phoneme duration prediction and a learnable upsampling layer $\theta_{\rm lu}$ for sequence expansion from phoneme level to frame level.

\paragraph{Duration Predictor}
The input to the duration predictor $\theta_{\rm dp}$ is phoneme hidden sequence $\mH_{n\times h}$ and the output is the estimated phoneme duration $\hat{\vd}_{n \times 1}$. The duration predictor $\theta_{\rm dp}$ consists of $3$ layers of one-dimensional convolution, with ReLU activation, layer normalization, and dropout between each layer.

\paragraph{Learnable Upsampling Layer}

The learnable upsampling layer $\theta_{\rm lu}$ takes phoneme duration $\vd$ as input and upsamples phoneme hidden sequence $\mH$ to frame-level sequence $\mO$~\citep{elias2021parallel}. First, we calculate the duration start and end matrices $\mS_{m\times n}$ and $\mE_{m\times n}$ by 
\begin{equation}
    \label{eq:learnable-upsampling-duration-matrics}
    \mS_{i,j}=i-\sum_{k=1}^{j-1}d_{k}, ~~~\mE_{i,j}=\sum_{k=1}^{j}d_{k} - i, 
\end{equation}
where $\mS_{i,j}$ indexes the $(i,j)$-th element in the matrix. We calculate the primary attention matrix $\mW_{m\times n \times q}$ and auxiliary context matrix $\mC_{m \times n \times p}$ following ~\cite{elias2021parallel}:
\begin{equation}
    \label{eq:learnable-upsampling-attention-matrix}
    \mW = \mathrm{Softmax}(\underset{10\rightarrow q}{\mathrm{MLP}}([\mS,\mE, \mathrm{Expand}(\mathrm{Conv1D}(\underset{}{\mathrm{Proj}}(\mH)))])),
\end{equation}
\begin{equation}
    \mC = \underset{10\rightarrow p}{\mathrm{MLP}}([\mS, \mE, \mathrm{Expand}(\mathrm{Conv1D}(\mathrm{Proj}(\mH)))]),
\end{equation}
where $\mathrm{Proj}(\cdot)$ represents one linear layer with input and output dimensions of $h$. $\mathrm{Conv1D}(\cdot)$ is one-dimensional convolution operation with layer normalization and Swish activation~\cite{ramachandran2017searching}. The input and output dimensions of $\mathrm{Conv1D}(\cdot)$ are $h$ and $8$. $\mathrm{Expand}(\cdot)$ means adding an extra dimension by repeating the input matrix by $m$ times. 
$[\cdot]$ stands for matrix concatenation along the hidden dimension, and gets a hidden dimension of $10 = 1+1+8$.
$\mathrm{MLP}(\cdot)$ is a two-layer full-connected network with Swish activations. The numbers underneath $\mathrm{MLP}$ denote the input and output hidden dimensions. We set $p=2$ and $q=4$. The $\mathrm{Softmax}(\cdot)$ operation is performed on the sequence time dimension. We calculate the frame-level hidden sequence output $\mO_{m\times d}$ with the following equation:

\begin{equation}
    \mO = \underset{qh \rightarrow h}{\mathrm{Proj}}(\mW \mH) + \underset{qp \rightarrow h}{\mathrm{Proj}}(\mathrm{Einsum}(\mW, \mC)), 
\end{equation}
where $\mathrm{Einsum}(\cdot)$ represents the einsum operation $(\mathrm{`qmn,mnp\rightarrow qmp\textrm'}, \mW, \mC)$. We first permute $\mW$ from $m\times n\times q$ to $q\times m\times n$ for computation, and after we get $\mW \mH$ with shape $q\times m\times h$ and $\mathrm{Einsum}(\mW, \mC)$ with shape $q\times m\times p$, we reshape them to $m\times qh$ and $m\times qp$ respectively for final projection to dimension $m \times h$. Finally, we map $\mO$ with a mean and variance linear layer to get the frame-level prior distribution parameter $\mu(y; \theta_{\rm pri})$ and $\sigma(y; \theta_{\rm pri})$, and get the prior distribution $p(z'|y; \theta_{\rm pri}) = \mathcal{N} (z'; \mu(y; \theta_{\rm pri}), \sigma(y; \theta_{\rm pri}))$.

Compared to simply repeating each phoneme hidden sequence with the predicted duration in a hard way, the learnable upsampling layer enables more flexible duration adjustment for each phoneme. Also, the learnable upsampling layer makes the phoneme to frame expansion differentiable, and thus can be jointly optimized with other modules in the TTS system.

\section{Alternative Formulation of Bidirectional Prior/Posterior}
\label{appendix_bpp}
We provide another formulation of the backward loss $\mathcal{L}_{\text{bwd}}$ in Equation~\ref{eqa_kl_backward} and forward loss $\mathcal{L}_{\text{fwd}}$ in Equation~\ref{eqa_kl_forward} by directly using KL loss to match two distributions. 

For the backward loss, we directly match the posterior $q(z|x;\phi)$ to the prior $p(z|y; \theta_{\text{pri}})$:
\begin{equation}
\begin{aligned}
\mathcal{L}_{\text{bwd}}(\phi,\theta_{\text{bpp}},\theta_{\text{pri}}) & =  KL[q(z|x;\phi)||p(z|y; \theta_{\text{pri}}))]  = \mathbb{E}_{z \sim q(z|x;\phi)} (\log q(z|x;\phi) - \log p(z|y; \theta_{\text{pri}})) \\
& = \mathbb{E}_{z \sim q(z|x;\phi)} (\log q(z|x;\phi) - \log p(f^{-1}(z;\theta_{\text{bpp}})|y; \theta_{\text{pri}})) \arrowvert \det \frac{\partial f^{-1}(z; \theta_{\text{bpp}})}{\partial z}  \arrowvert), \\
\end{aligned}
\label{eqa_kl_backward_1}
\end{equation}
where $f^{-1}(z; \theta_{\text{bpp}}) = z'$, and $p(z|y; \theta_{\text{pri}})) = p(f^{-1}(z;\theta_{\text{bpp}})|y; \theta_{\text{pri}})) \arrowvert \det \frac{\partial f^{-1}(z; \theta_{\text{bpp}})}{\partial z}  \arrowvert)$ according to the change of variable rule. 

For the forward loss, we directly match the prior $p(z'|y; \theta_{\text{pri}})$ to the posterior $q(z'|x;\phi)$:
\begin{equation}
\begin{aligned}
\mathcal{L}_{\text{fwd}}(\phi,\theta_{\text{bpp}}, \theta_{\text{pri}}) & = KL[p(z'|y; \theta_{\text{pri}})||q(z'|x;\phi)] = \mathbb{E}_{z' \sim p(z'|y; \theta_{\text{pri}})} (\log p(z'|y; \theta_{\text{pri}}) - \log q(z'|x;\phi)) \\
& = \mathbb{E}_{z' \sim p(z'|y; \theta_{\text{pri}})} (\log p(z'|y; \theta_{\text{pri}}) - \log q(f(z';\theta_{\text{bpp}})|x; \phi) \arrowvert \det \frac{\partial f(z';\theta_{\text{bpp}})}{\partial z'}  \arrowvert),
\end{aligned}
\label{eqa_kl_forward_1}
\end{equation}
where $f(z'; \theta_{\text{bpp}}) = z$, and $q(z'|x;\phi)) = q(f(z';\theta_{\text{bpp}})|x; \phi) \arrowvert \det \frac{\partial f(z';\theta_{\text{bpp}})}{\partial z'}  \arrowvert)$ according to the change of variable rule.

\section{Soft Dynamic Time Warping in KL loss}
\label{appendix_soft_dtw}

Since the frame-level prior distribution $p(z'|y;\theta_{\rm pri})$ usually has different lengths from the ground-truth speech frames, the standard KL loss cannot be applied. Therefore, we use a soft dynamic time warping (Soft-DTW) of KL loss for $\mathcal{L}_{\rm bwd}$ and $\mathcal{L}_{\rm fwd}$ to circumvent this mismatch.

The Soft-DTW version of the KL loss for $\mathcal{L}_{\rm bwd}$ can be obtained by recursive calculation:
\begin{equation}
    r_{i,j} = {\min}^{\gamma} \left\{
    \begin{aligned}
        & r_{i-1,j} + KL[q(z'_{i-1}|x;\phi,\theta_{\rm bpp}) || p(z'_{j}|y;\theta_{\rm pri})] + \mathrm{warp} \\
        & r_{i, j-1} + KL[q(z'_{i}|x;\phi,\theta_{\rm bpp}) || p(z'_{j-1}|y;\theta_{\rm pri})] + \mathrm{warp} \\
        & r_{i-1, j-1} + KL[q(z'_{i-1}|x;\phi,\theta_{\rm bpp}) || p(z'_{j-1}|y;\theta_{\rm pri})]
    \end{aligned}
\right.,
\end{equation}
where $r_{i,j}$ is the KL divergence loss between the simplified posterior $q(z'|x;\phi,\theta_{\rm bpp})$ from frame $1$ to frame $i$ and the prior $p(z'|y; \theta_{\rm pri})$ from frame $1$ to frame $j$ with the best alignment. $KL[q(z'_*|x;\phi,\theta_{\rm bpp})|| p(z'_*|y;\theta_{\rm pri})]$ is defined in Equation~\ref{eqa_kl_backward}. $\min^{\gamma}$ a soft-min operator, which is defined as $\min^{\gamma} (a_1,...,a_n) = - \gamma \log \Sigma_i e^{-\frac{a_i}{\gamma}}$ and $\gamma=0.01$. $\mathrm{warp}$ is a warp penalty for not choosing the diagonal path and is set as $0.07$. $q(z'_{i}|x;\phi,\theta_{\rm bpp})$ is the $i$-th frame of the simplified posterior, and $p(z'_{j}|y;\theta_{\rm pri})$ is the $j$-th frame of the prior.

The Soft-DTW version of KL loss for $\mathcal{L}_{\rm fwd}$ is similar to that of $\mathcal{L}_{\rm bwd}$, which can be defined as:
\begin{equation}
    r_{i,j} = {\min}^{\gamma} \left\{
    \begin{aligned}
        & r_{i-1,j} + KL[p(z_{i-1}|y;\theta_{\rm pri},\theta_{\rm bpp}) || q(z_{j}|x;\phi)] + \mathrm{warp} \\
        & r_{i, j-1} + KL[p(z_{i}|y;\theta_{\rm pri},\theta_{\rm bpp}) || q(z_{j-1}|x;\phi)] + \mathrm{warp} \\
        & r_{i-1, j-1} + KL[p(z_{i-1}|y;\theta_{\rm pri},\theta_{\rm bpp}) || q(z_{j-1}|x;\phi)]
    \end{aligned}
\right.,
\end{equation}
where $r_{i,j}$ is the KL divergence loss between the enhanced prior $p(z|y;\theta_{\rm pri},\theta_{\rm bpp})$ from frame $1$ to frame $i$ and the posterior $q(z|x;\phi)$ from frame $1$ to frame $j$ with the best alignment. $KL[p(z_*|y;\theta_{\rm pri},\theta_{\rm bpp}) || q(z_*|x;\phi)]$ is defined in Equation~\ref{eqa_kl_forward}. $p(z_{i}|y;\theta_{\rm pri},\theta_{\rm bpp})$ is the $i$-th frame of the enhanced prior, and $q(z_{j}|x;\phi)$ is the j-th frame of the posterior.

\section{Waveform Decoder Loss}
\label{appendix_wav_dec_loss}

Instead of using negative log-likelihood loss in waveform reconstruction and prediction in Equation~\ref{eqa_vae_rec} and~\ref{eqa_fully_e2e}, we use GAN loss, feature mapping loss, and mel-spectrogram loss as used in ~\citep{kong2020hifi}.

\paragraph{GAN Loss}
The GAN loss follows LS-GAN~\citep{mao2017least}, which is defined as follows. The generator is trained to minimize the loss function while the discriminator is train to maximize it:
\begin{equation}
    \mathbb{E}_x[(D(x) - 1)^2] + \mathbb{E}_z[D(G(z))^2]
\end{equation}
where is $x$ the ground-truth waveform and $z$ is the input of waveform decoder. We follow~\citep{kim2021conditional} for the design of discriminators.

\paragraph{Feature Mapping Loss}
The feature mapping loss consists of the L1 distance between real samples and fake samples in terms of the intermediate feature in each layer of the discriminator, which can be formulated as:
\begin{equation}
    \mathbb{E}_{(x,z)}[\sum_{l} \frac{1}{N_l} ||D^{l}(x) - D^{l}(G(z))||_1]
\end{equation}
where $l$ is the layer index in discriminator, $D^l(\cdot)$ and $N_l$ are the features and the number of features in the $l$-th layer of the discriminator, respectively.

\paragraph{Mel-Spectrogram Loss}
The mel-spectrogram loss is L1 distance between the mel-spectrogram of ground-truth waveform and that of generated waveform, which can be defined as:
\begin{equation}
    \mathbb{E}_{(x,z)} = ||S(x) - S(G(z))||_1
\end{equation}
where $S(\cdot)$ is the function that converts the waveform into corresponding mel-spectrogram.

\section{Training Details of \myname{}}
\label{appendix_training_details}

\paragraph{Phoneme Pre-training}
We pre-train our phoneme encoder on $200$M phoneme sequences, which is converted from text with grapheme-to-phoneme conversion. The size of the phoneme dictionary is $182$. We learn the sup-phoneme using Byte-Pair Encoding (BPE)~\citep{sennrich2015neural} with a sup-phoneme dictionary size of $30,088$. We conduct the pre-training on $8$ NVIDIA A100 GPUs with 80G memory (we only use A100 for phoneme pre-training, and use V100 for the remaining training of \myname{}), with a total batch size of $1,024$ sentences for $120$k training steps. The mask ratio for sup-phoneme is $15\%$.  

\paragraph{Duration Predictor}
In the warmup stage (the first $1$k epochs), we obtain the duration label to train the duration predictor to speed up the convergence of differentiable durator. We can choose any tools to provide duration label, such as Montreal forced alignment~\citep{mcauliffe2017montreal}. Here we choose monotonic alignment search~(MAS)~\cite{kim2020glow}, which estimates the optimal alignment between the phoneme prior distribution $p(z'|y; \theta_{\rm pho}) = \mathcal{N} (z'; \mu(y; \theta_{\rm pho}), \sigma(y; \theta_{\rm pho}))$ and simplified frame-level posterior $q(z'|x;\phi,\theta_{\rm bpp})$, where $\mu(y; \theta_{\rm pho}), \sigma(y; \theta_{\rm pri})$ are the mean and variance parameters obtained from the phoneme hidden sequence by two linear layers. The monotonic and non-skipping constraints of MAS provide the inductive bias that human read words in orders without skipping. The optimal alignment search result $A$ can be formulated as
\begin{equation}
    A = \underset{A}{\mathrm{arg}~\mathrm{max}}~\Sigma_{i=1}^{m}\mathcal{N}(z'_i; \mu(y; \theta_{\rm pho})_{A(i)}, \sigma(y; \theta_{\rm pho})_{A(i)}),
\end{equation}
where $A(i)$ denotes the aligned phoneme index of the $i$-th frame $z^{\prime}_{i}$ from $q(z'|x;\phi,\theta_{\rm bpp})$. 
We search the alignment result using dynamic programming. Let $Q_{i,j}$ denote the probability of $z^{\prime}_{i}$ belongs to the prior distribution of the $j$-th phoneme, then we can formulate $Q_{i,j}$ recursively with $Q_{i-1,j-1}$ and $Q_{i,j-1}$ with the following equation:

\begin{equation}
    Q_{i,j} = \mathrm{max}(Q_{i-1,j-1}, Q_{i-1,j}) + \mathrm{log}~\mathcal{N}(z'_i; \mu(y; \theta_{\rm pho})_j, \sigma(y; \theta_{\rm pho})_j).
\end{equation}
We calculate all the $Q_{i,j}$ from $i=0,j=0$ to $i=m,j=n$. Since the best alignment path is determined by the highest $Q$ value, we utilize all the cached $Q$ value to backtrack from $Q_{m, n}$ to $Q_{0, 0}$ for the most probable alignment $A$.

Note that in the warmup training stage, the duration $\vd$ comes from MAS. After the warmup stage, the input duration comes from the duration predictor $\hat{\vd}$. During the whole training process, we apply gradient stop operation on the input of duration predictor.

\paragraph{Bidirectional Prior/Posterior}
For the two loss terms $\mathcal{L}_{\text{bwd}}$ and $\mathcal{L}_{\text{fwd}}$ in bidirectional prior/posterior module, we only use $\mathcal{L}_{\text{bwd}}$ during the warmup stage to learn a reasonable prior distribution, and then add $\mathcal{L}_{\text{fwd}}$ to the loss function for bidirectional optimization after the warmup stage.

\paragraph{VAE with Memory}
In the warmup stage, we do not use the memory bank in VAE training, i.e., $z \sim q(z|x;\phi)$ is directly taken as the input of the waveform decoder. After the warmup stage, we initialize the memory banks $M$ as follows: we first get the posterior $z \sim q(z|x;\phi)$ of each frame of the utterances in the training set, and then conduct K-means clustering on these $z$ to get $1$K clusters, and use the cluster center to initialize the memory bank $M$. After the initialization, we jointly train the memory mechanism with the whole TTS system.

In the tuning stage (the last $2$k epochs), we only use $\mathcal{L}_{\text{e2e}}$ to tune the model. We freeze the parameters of posterior encoder, waveform decoder, phoneme encoder, and bidirectional prior/posterior, and only update the durator for fully end-to-end duration optimization.

\section{Hyper-Parameters of \myname{}}
\label{appendix_hyperpara}
The hyper-parameters of \myname{} are listed in Table~\ref{tab_hyperpara}.

\begin{table}[h!]
\small
    \centering
    \caption{Hyper-parameters of \myname{}.}
    \begin{tabular}{llc}
    \toprule
    Module & Hyper-Parameter     &   Value  \\
    \midrule
    \multirow{7}{*}{Phoneme Encoder $\theta_{\rm pho}$} & Phoneme Encoder Embedding Dimension &  $192$     \\
    &Phoneme Encoder Blocks  & $6$ \\
    &Phoneme Encoder Multi-Head Attention Hidden Dimension & $192$ \\
    &Phoneme Encoder Multi-Head Attention Heads & $2$ \\
    &Phoneme Encoder Conv Kernel Size & $3$ \\
    &Phoneme Encoder Conv Filter Size & $768$ \\
    &Phoneme Encoder Dropout & $0.1$ \\
    \midrule
    \multirow{5}{*}{Durator $\theta_{\rm dur}$}&Duration Predictor Kernel Size & $3$ \\
    &Duration Predictor Filter Size & $192$ \\
    &Duration Predictor Dropout & $0.5$ \\
    &Upsampling Layer Kernel Size & $3$ \\
    &Upsampling Layer Filter Size & $8$ \\
    \midrule
    \multirow{5}{*}{Prior/Posterior $\theta_{\rm bpp}$}&Flow Model Affine Coupling Layers & $4$ \\
    &Flow Model Affine Coupling Dilation & $1$ \\
    &Flow Model Affine Coupling Kernel Size & $5$ \\
    &Flow Model Affine Coupling Filter Size & $192$ \\
    &Flow Model Affine Coupling WaveNet Layers & $4$ \\
    \midrule
    \multirow{6}{*}{Waveform Decoder $\theta_{\rm dec}$}&Waveform Decoder ConvBlocks & $4$ \\
    &Waveform Decoder ConvBlock Hidden & $[256, 128, 64, 32]$ \\
    &Waveform Decoder ConvBlock Upsampling Ratio & $[8, 8, 2, 2]$ \\
    &Waveform Decoder ConvLayers & $3$ \\
    &Waveform Decoder ConvLayer Kernel Size & $[3,7,11]$ \\
    &Waveform Decoder Conv Dilation & $[1,3,5]$ \\
    &Memory Banks Size & $1000$ \\
    &Memory Banks Hidden Dimension & $192$ \\
    &Memory Banks Attention Heads & $2$ \\
    \midrule
    \midrule
    \multirow{5}{*}{Posterior Encoder $\phi$}&Posterior Encoder WaveNet Layers & $16$ \\
    &Posterior Encoder Dilation & $1$  \\
    &Posterior Encoder Conv Kernel Size & $5$ \\
    &Posterior Encoder Conv Filter Size & $192$ \\
    \midrule
    Discriminator $D$ & Multi-Period Discriminator Periods & $[1, 2, 3, 5, 7, 11]$ \\
    \bottomrule
    \end{tabular}
    \label{tab_hyperpara}
\end{table}

The number of model parameters for $\theta_{\rm pho}$, $\theta_{\rm dur}$, $\theta_{\rm bpp}$, and $\theta_{\rm dec}$ is $28.7$M, for the posterior encoder $\phi$ is $7.2$M, and for the discriminators is $46.7$M. Note that only $\theta_{\rm pho}$, $\theta_{\rm dur}$, $\theta_{\rm bpp}$, and $\theta_{\rm dec}$ with $28.7$M model parameters are used in inference.

\end{document}